\documentclass[preprint,3p,12pt]{elsarticle}

\usepackage{amssymb}
\usepackage{amsmath}
\usepackage{graphicx}
\usepackage{xcolor}
\usepackage[colorlinks=true,pdfa=true
]{hyperref}

\journal{PoS: 38th ICRC 2023}

\begin{document}

\begin{frontmatter}



\title{Spine-sheath jet model for low-luminosity AGNs}


\author[a]{Margot Boughelilba\corref{cor1}}
\cortext[cor1]{Corresponding author}
\ead{margot.boughelilba@uibk.ac.at}

\author[a]{Anita Reimer}
\ead{anita.reimer@uibk.ac.at}

\author[b,c]{Lukas Merten}
\ead{lukas.merten@rub.de}

\author[d]{Jon Paul Lundquist}
\ead{jplundquist@gmail.com}

\affiliation[a]{organization={Universität Innsbruck, Institut für Astro- und Teilchenphysik},
            city={Innsbruck},
            postcode={6020}, 
            country={Austria}}
\affiliation[b]{organization={Theoretical Physics IV, Plasma Astroparticle Physics, Faculty for Physics and Astronomy, Ruhr University Bochum},
            city={Bochum},
            postcode={44780}, 
            country={Germany}}
\affiliation[c]{organization={Ruhr Astroparticle and Plasma Physics Center (RAPP Center)},
            city={Bochum},
            postcode={44780}, 
            country={Germany}}
\affiliation[d]{organization={Center for Astrophysics and Cosmology (CAC), University of Nova Gorica},
            city={Nova Gorica},
            postcode={SI-5000}, 
            country={Slovenia}}

\begin{abstract}

In several jetted AGNs, structured jets have been observed. In particular spine-sheath configurations where the jet is radially divided into two or more zones of different flow velocities. 

We present a model based on the particle and radiation transport code CR-ENTREES. Here, interaction rates and secondary particle and photon yields are pre-calculated by Monte Carlo event generators or semi-analytical approximations. These are then used to create transition matrices, that describe how each particle spectrum evolves with time. This code allows for arbitrary injection of primary particles, and the possibility to choose which interaction to include (photo-meson production, Bethe-Heitler pair-production, inverse-Compton scattering, $\gamma$-$\gamma$ pair production, decay of all unstable particles, synchrotron radiation --- from electrons, protons, and all relevant secondaries before their respective decays --- and particle escape). 

In addition to the particle and radiation interactions taking place in each homogeneous zone, we implement the feedback between the two zones having different bulk velocities. The main mechanism at play when particles cross the boundary between the two zones is shear acceleration. We follow a microscopic description of this acceleration process to create a corresponding transition matrix and include it in our numerical setup. Furthermore, each zone's radiation field can be used as an external target photon field for the other zone's particle interactions. We present here the first results of the effect of a two-zone spine-sheath jet, by applying this model to typical low-luminosity AGNs.

\end{abstract}

\end{frontmatter}


\section{Introduction}

Structured jets, and in particular spine-sheath configurations, are naturally formed in astrophysical environments because of density and velocity gradients. Observations hinting towards such transversal structures have been reported in the case of Active Galactic Nuclei (AGNs), e.g. for the blazar Mkn 501 \citep{Spine_Sheath_Mkn501} or for the radio-galaxy M87 \citep{Spine_Sheath_M87}. Furthermore, several relativistic and magneto-hydrodynamic simulations of two-component AGN jets have demonstrated that the presence of a sheath stabilizes the jet (see, e.g. \citep{3D_RMHD,GRMHD,transverse_stability}).

Large-scale jets of AGNs are particularly interesting to study the impact a spine-sheath configuration can have. Detection of non-thermal X-ray emission along the extended, kpc-scale jet indicates the presence of (ultra-)relativistic particles \citep{Xrays_large_scale, Xrays_large_scale_review}. In the classically preferred scenario, this emission is attributed to electron synchrotron radiation, implying in this case very large Lorentz factors $\gamma \sim 10^8$ \citep{lorentz_factor_electrons}. Due to the typically very short cooling length of electrons with such energies, there must be a (re-)acceleration process occurring continuously or distributed along the jet, to keep the electrons energized. Stochastic shear acceleration can be one of these continuous (re-)acceleration processes \citep{Spine_Sheath_synch, FR0s}.

When energetic particles cross the shear layer between the spine and the sheath, they can be accelerated to higher energies. This mechanism has been shown to be an efficient re-acceleration process in specific environments. Similarly to Fermi acceleration, shear acceleration is due to the fact that particles can gain energy by scattering off (small-scale) magnetic field inhomogeneities moving with different local velocities (see \citep{2019Galax...7...78R} for a review). 

In this framework, this re-acceleration can also facilitate the acceleration of cosmic rays (CRs) up to very high energies (see e.g. \citep{FR0s,UHECR_ostrowski,CR_Webb,UHECR_kimura}).

In this work, we describe the implementation of shear acceleration into the CR-ENTREES -- Cosmic-Ray ENergy TRansport in timE-Evolving astrophysical Settings -- code (see \citep{Anita_CR_ENTREES}, this conference). CR-ENTREES solves the coupled time- and energy-dependent kinetic equations for protons, pions, muons, electrons, positrons, photons and neutrinos in a one-zone setup, with a free choice of particle and photon injection. The radiation-dominated, magnetized astrophysical environment is evolving in time and all relevant interactions, as well as particle and photon escape, are considered. Particle and photon interactions are pre-calculated using event generators. We use the matrix multiplication method for transporting the radiation and particle energy, including the non-linear feedback from radiation and particle re-injected into the simulation chain after interacting.

\section{Two-zone models in CR ENTREES}

We use the CR-ENTREES code to compute the densities of the particles in the comoving jet frame (see \citep{Anita_CR_ENTREES}). CR-ENTREES -- Cosmic-Ray
ENergy TRansport in timE-Evolving astrophysical Settings -- is fully time-dependent and its modular structure allows us to implement new features, such as shear acceleration and spatial diffusion, in the case of two-zone jets.
\subsection{Shear Acceleration}
\begin{figure}[ht]
    \centering
    \includegraphics[width = \textwidth]{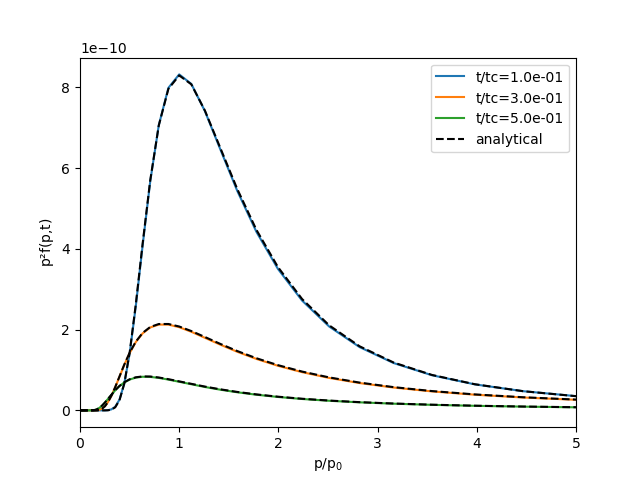}
    \caption{Comparison between the analytical solution Eq. \ref{anaylitical_solution} (black lines) and the numerical solution (yellow, pink and cyan lines) of the Fokker-Planck Eq. \ref{dif_eq2} using CR-ENTREES at times $t = 0.1,\,  0.3,\,  0.5 \, t_c$ respectively, where $t_c = (\alpha^2 \Gamma \tau_0 p_0^\alpha)^{-1}$, for $\alpha = 1$.}
    \label{fig:anaytical}
\end{figure}

The effect of shear acceleration on the particles' distribution function is described by a Fokker-Planck differential equation, assuming no particle escape, and no energy losses or further acceleration:
\begin{equation}
    \frac{\partial f}{\partial t} = \frac{1}{p^2}\frac{\partial}{\partial p}\left( p^2 D \frac{\partial f}{\partial p} \right) + Q(p,t) \label{dif_eq}
\end{equation}
where $f(p,t)$ is the momentum-space particle distribution, $p$ and $t$ are the momentum and time respectively, $D$ denotes the momentum space diffusion coefficient and $Q(p,t)$ is the source term. 
For an impulsive mono-energetic source term $Q(p,t) = Q_0\delta(p-p_0)\delta(t)$, this equation has been shown to have an analytical solution \citep{Rieger_analytical}
\begin{equation}
    f(p,t) = \frac{Q_0 p_0^{(\alpha+1)}}{|\alpha| \Gamma \tau_0 t}\left(\frac{p_0}{p}\right)^{(3+\alpha)/2}\, \exp{\left(-\frac{p^{-\alpha} + p_0^{-\alpha}}{\alpha^2\Gamma\tau_0 t}\right)}\, I_{|1+3/\alpha|}\left[\frac{2}{\alpha^2 \Gamma \tau_0 p_0^\alpha t}\left(\frac{p}{p_0}\right)^{(-\alpha/2)}\right], \label{anaylitical_solution}
\end{equation}
where $I_{\nu}(z)$ is the modified Bessel function of the first kind \citep{math_functions},
we assume that the local scattering time $\tau$ is a power-law function of momentum such that $\tau = \tau_0 p^\alpha$ and $\Gamma$ is the shear flow coefficient and depends on the velocity gradient in the shear layer. For a gradual non-relativistic two-dimensional shear flow with a velocity profile $\Vec{u} = u_z(r)\Vec{e_z}$, where the z-direction is along the jet axis and the r-direction is the transverse radial direction, the shear flow coefficient is given by $\Gamma = (1/15)(\partial u_z/\partial r)^2$ (in the case of a relativistic flow, one can replace $\Gamma \rightarrow \Gamma = (1/15)(\Gamma_j^2\partial u_z/\partial r)^2$, where $\Gamma_j$ is the bulk Lorentz factor of the flow, \citep{2019Galax...7...78R}). The momentum space diffusion coefficient is then written as $D = \Gamma p^2 \tau$ \citep{Rieger_analytical}.
Eq. \ref{anaylitical_solution} is however, valid only at large times, and we turned to numerical methods to solve the differential equation.
Following \citep{Park_Petrosian} we rewrite Eq. \ref{dif_eq} as:

\begin{equation}
    \frac{\partial f}{\partial t} = \frac{1}{A(x)}\frac{\partial}{\partial x}\left[C(x) \frac{\partial f}{\partial x} + B(x)f\right] - \frac{u}{T(x)} + Q(x,t) \label{dif_eq2}.
\end{equation}
Where $x$ represents either the momentum or the energy $E$, $A$ is the phase factor, such that 
\begin{equation}
    A(x) = \begin{cases}
    1 & \text{if $x=E$ or $\gamma$}\\
    4\pi x^2 & \text{if $x=p$}
  \end{cases}
\end{equation}
With these definitions, $f(x,t)A(x)dx$ is the number of particles in the interval $x$ and $x+dx$ at a time t. 
In this way, Eq. \ref{dif_eq2}, combined with a no-flux boundary condition can be transformed into a tri-diagonal system of linear equations and solved with a Gaussian elimination algorithm. 
The terms $B(x)$, $C(x)$, $T(x)$ and $Q(x)$ belong to the advection, diffusion, escape and source terms respectively. Comparing Eq. \ref{dif_eq} to Eq. \ref{dif_eq2} one yields $A(x) = 4\pi p^2$, $B(x) = 0$, $C(x) = 4\pi D p^2$, $T(x) \rightarrow \infty$ and $Q(x,t) = Q(p,t)$, if we use $x=p$.
If instead, we choose $x = \gamma$, assuming $p \approx \gamma m c$ for relativistic particles and assuming an isotropic particle distribution, then $n(\gamma,t)d\gamma = 4\pi p^2 f(p,t)dp$. Re-writing Eq. \ref{dif_eq} in the form of Eq. \ref{dif_eq2} then yields $A(x) = 1$, $B(x) = -(2/\gamma)D$, $C(x) = D$, $T(x) \rightarrow \infty$ and $Q(x,t) = Q(\gamma,t)$.
Following \citep{Park_Petrosian} we use the fully implicit method (see equations (25)-(28) in \citep{Park_Petrosian}) to solve equation \ref{dif_eq2}. In our case, we find satisfactory results with this simple method (see Figure \ref{fig:anaytical}), using $x = \gamma$.

\begin{figure}[ht]
    \centering
    \includegraphics[width = \textwidth]{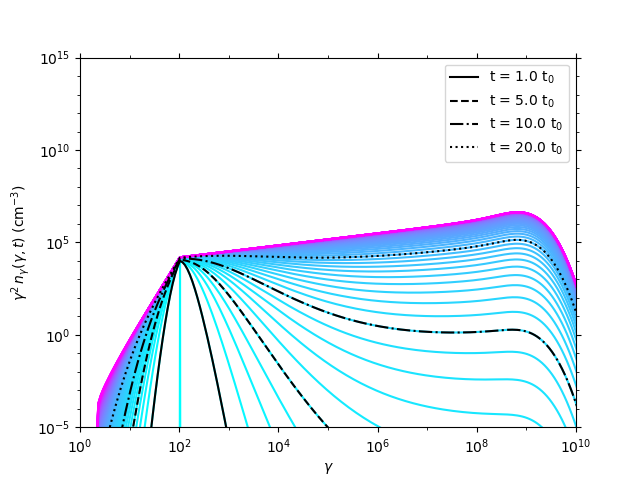}
    \caption{Evolution of the electron spectrum, for a continuous mono-energetic injection of electrons at $\gamma_0 = 100$. The particles are accelerated via shear acceleration and experience synchrotron losses at the same time. The black lines highlight snapshots of the simulation, in order to compare with the results of \citep{Spine_Sheath_synch}.}
    \label{fig:synch}
\end{figure}

We further tested the implementation of shear acceleration in CR-ENTREES by running a simulation where electrons experience both acceleration and synchrotron losses. The parameters are the same as in \citep{Spine_Sheath_synch}: the magnetic field strength $B = 3 \mu\mathrm{G}$, the electrons are injected continuously with an initial Lorentz factor of $\gamma_0 = 100$, the shear layer has a width $\Delta L = 10^{19}\, \mathrm{cm}$ and is located in the jet between $9\times 10^{19}\, \mathrm{cm}$ and $10^{20}\, \mathrm{cm}$. The velocity profile of the jet is linear and the bulk Lorentz factor goes from $\Gamma_\mathrm{j,in} = 1.1$ to
 $\Gamma_\mathrm{j,out} = 1.0$. The other relevant parameters for shear acceleration are set to $q = 2 - \alpha = 1.67$, $\xi = 0.1$ and $\Lambda_\mathrm{max} = 10^{18}$, where the latter two are related to the mean free path of the particles such that $\lambda = \xi^{-1}r_g\left(\frac{r_g}{\Lambda_\mathrm{max}}\right)^{(1-q)}$, with $r_g$ the Larmor radius of the particle, $\xi$ the ratio between the turbulent and regular magnetic field energy densities and $\Lambda_\mathrm{max}$ the longest interacting wavelength of the turbulence. Figure \ref{fig:synch} shows the time evolution of the electron spectrum. The cutoff at high energies is when the acceleration timescale and the loss timescale are comparable. We highlighted five timesteps to compare with the results of \citep{Spine_Sheath_synch}. 

\subsection{External target fields}
\label{EC}
Considering two-zone models gives a new source of target photon fields for the interactions occurring in the jet. Indeed, the photons emitted by the spine (sheath) can serve as external targets for the particles in the sheath (spine). Because both layers move with different velocities, the external emission of one layer is seen boosted in the comoving frame of the other layer. The transformation of the spectral energy density between the external frame (unprimed quantities) and the comoving frame (primed quantities) is given by
\begin{equation}
    u(E',\Omega') = \frac{u(E,\Omega)}{\Gamma_\mathrm{rel}^2(1 - \mu' \beta_\mathrm{rel})},
\end{equation}
where $\mu'$ is photon direction cosine, $\beta_\mathrm{rel}c = c(\beta_\mathrm{j,in} - \beta_\mathrm{j,out})/(1 - \beta_\mathrm{j,in}\,\beta_\mathrm{j,out})$ is the relative bulk velocity between the two frames and $\Gamma_\mathrm{rel} = (1-\beta_\mathrm{rel}^2)^{-1/2}$ is the corresponding relative bulk Lorentz factor. The external photon energy is related to the comoving energy by $E = \Gamma_\mathrm{rel}E'(1+\mu'\beta_\mathrm{rel})$ and $\mu = (\mu' +\beta_\mathrm{rel})/(1+\mu'\beta_\mathrm{rel})$.

For a mono-energetic isotropic radiation field $u(E,\Omega) = u_0\,\delta(E-E_0)/(4\pi)$ we get 
\begin{equation}
    u'(E') = u_0 \Gamma_\mathrm{rel}^2(1 - \beta_\mathrm{rel}^2/3)\delta(E'-E'_0).
\end{equation}
In the comoving frame, the external emission is not isotropic. However, $u'(E')$ peaks at $E' = E_0/\Gamma_\mathrm{rel}/(1-\beta_\mathrm{rel})$ (i.e. for $\mu' = -1)$, and we make the approximation that all the external emission in the comoving frame peaks at this energy and hence is also mono-energetic, in order to implement external photon fields in CR-ENTREES. 

\subsection{Spatial diffusion}

For shear acceleration to occur in the jet, particles need to cross the shear layer. This can happen because particles escape their initial emission region, or when they diffuse due to a density gradient ($\Delta n/\Delta L$). We calculate the number $N$ of particles diffusing through the shear layer surface $S$ within a time $\Delta t$ with Fick's law:
\begin{equation}
	N = \kappa(\gamma) \, \frac{\Delta n}{\Delta L} \, \Delta t\, S 
\end{equation}
where $\kappa(\gamma) = \tau\,c^2/3$ is the spatial diffusion coefficient. Only these particles will experience shear acceleration as they pass through the shear layer. 

\section{Example: a typical low-luminosity AGN}

\begin{figure}[h]
    \centering
    \includegraphics[width = \textwidth]{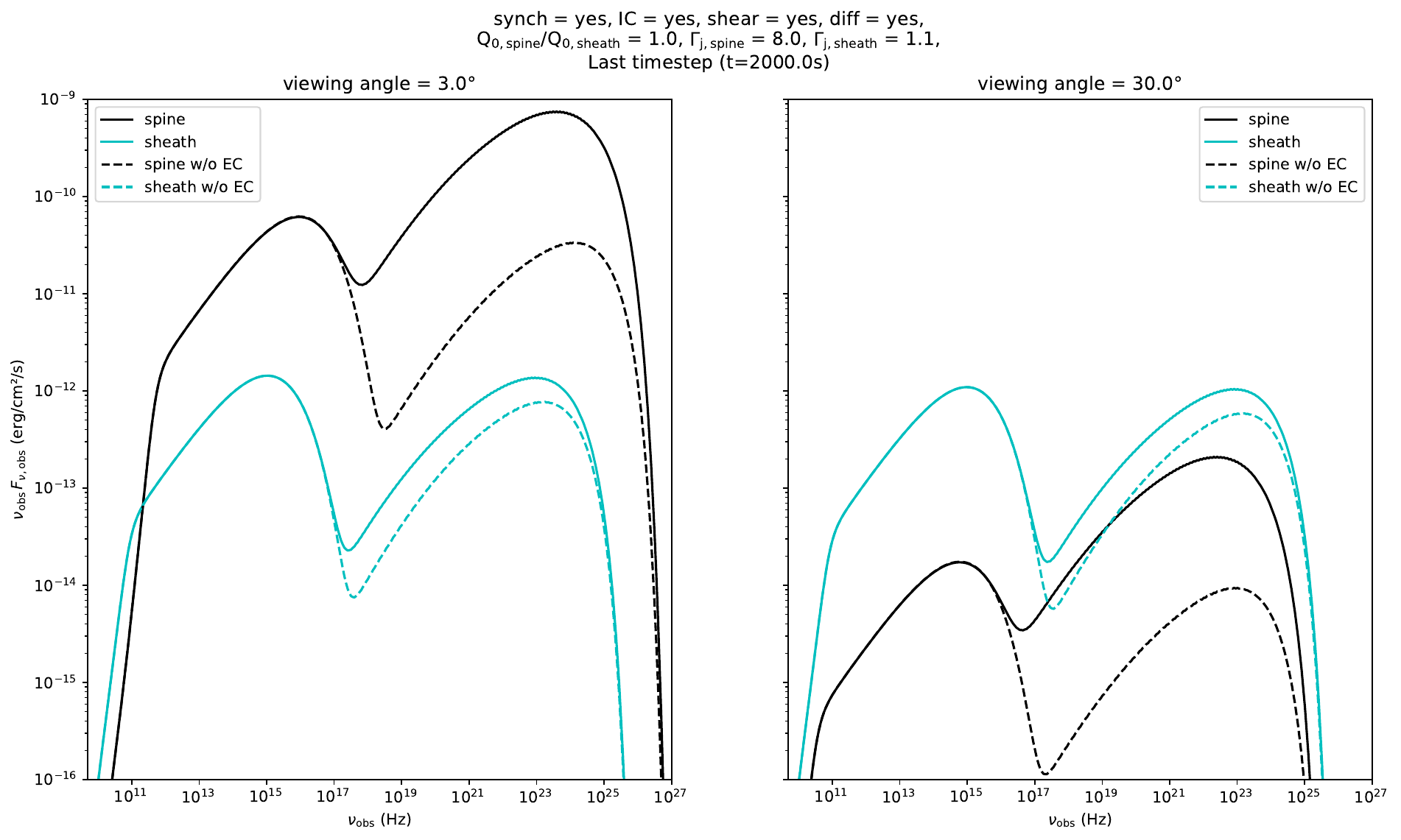}
    \caption{Observed photon spectral energy distributions, solid lines show the simulation taking into account the additional external target field, while dashed lines show the same simulation without this effect. Left: Blazar-like case, with a jet inclination of $3^\circ$. Right: Radio-galaxy-like case, with a jet inclination of $30^\circ$.}
    \label{fig:RG_with_ext}
\end{figure}

We show the results of two simulations of cylindrical jets where all the features described above are implemented. Electrons are injected at $t=0$ with a power-law spectrum of index $p_\mathrm{el}=2$, in the spine and the layer, with an energy density ratio of $1$. They suffer from synchrotron losses and further interact with the synchrotron photons (Synchrotron Self-Compton case) and experience shear acceleration in between the two layers. Furthermore, as discussed in Section \ref{EC}, photons emitted in the other layer also serve as target field for the electrons in the comoving frame (External Compton case). Figure \ref{fig:RG_with_ext} shows the observed photon spectrum, where both External Compton and Synchrotron Self-Compton interactions are taken into account, with solid lines. The jet is assumed to be observed with an angle $\theta_\mathrm{j} = 3^\circ$ or $\theta_\mathrm{j} = 30^\circ$, representing a typical blazar or radio galaxy respectively. The spine has a bulk Lorentz factor of $\Gamma_\mathrm{j,in} = 8$ and the sheath $\Gamma_\mathrm{j,out} = 1.1$. The shear layer has a width of $5\times 10^{15}\,\mathrm{cm}$, the spine has a radius of $3\times 10^{16}\,\mathrm{cm}$ and the layer spans from $3.5\times 10^{16}\,\mathrm{cm}$ to $7.5\times 10^{16}\,\mathrm{cm}$. The length of the cylinder is fixed to $7.5\times 10^{17}\,\mathrm{cm}$ We also show the same simulation, without the External Compton interaction, to highlight the importance of such process, represented in dashed lines.

\section{Conclusion}

We have implemented shear acceleration in the CR-ENTREES code and tested this process with simple cases. The two-zone model requires further mechanisms to be taken into account like External Compton interaction between the layers' frames, or spatial diffusion due to particle density gradients. These combined interactions are tested in the simulation shown in Figure \ref{fig:RG_with_ext}. The importance of External Compton is visible, especially for the spine, which has a larger bulk Lorentz factor than the sheath. This figure also shows the difference between a blazar jet, seen face-on, compared to a radio galaxy jet, where the jet is more inclined. Indeed, in the latter case, the sheath is dominant, even though it is very mildly relativistic, while the relativistic spine is of less importance.

\section*{Acknowledgments}

Financial support was received from the Austrian Science Fund (FWF) under grant agreement number I 4144-N27 and the Slovenian Research Agency-ARRS (project no. N1-0111). MB has for this project received funding from the European Union’s Horizon 2020 research and innovation program under the Marie Sklodowska-Curie grant agreement No 847476. The views and opinions expressed herein do not necessarily reflect those of the European Commission.




\bibliographystyle{elsarticle-num-names} 
\bibliography{main}





\end{document}